\documentclass[doublecol]{epl2}
\usepackage{amsmath}
\usepackage{url}
\usepackage{caption}
\usepackage{subcaption}
\usepackage{grffile}
\usepackage[latin2]{inputenc}
\usepackage{microtype}

\title{Symmetry detection of auxetic behaviour in 2D frameworks}

\author{%
  H.\ Mitschke$^*$\inst{1}\and%
  G.E.\ Schr\"oder-Turk\inst{1}\and%
  K.\ Mecke\inst{1}\and%
  P.W.\ Fowler\inst{2}\and%
  S.D.\ Guest\inst{3}%
}
\shortauthor{H.\ Mitschke \etal{}}

\institute{%
  \inst{1} Theoretische Physik I, Friedr.-Alex.\
           Universit\"at Erlangen-N\"urnberg, Staudtstr.\ 7, 91058 Erlangen, Germany\\%
  \inst{2} Department of Chemistry, University of Sheffield, Sheffield S3 7HF,
           UK\\%
  \inst{3} Department of Engineering, University of Cambridge, Trumpington
           Street, Cambridge CB2 1PZ, UK%
}

\pacs{62.20.dj}{Poisson's ratio}

\abstract{A symmetry-extended Maxwell treatment of the net mobility of
periodic bar-and-joint frameworks is used to derive a sufficient condition
for auxetic behaviour of a 2D material.  The type of auxetic behaviour that
can be detected by symmetry has Poisson's ratio $-1$, with equal
expansion/contraction in all directions, and is here termed
\emph{equiauxetic}.  A framework may have a symmetry-detectable equiauxetic
mechanism if it belongs to a plane group that includes rotational axes of
order $n = 6$, $4$, or $3$.  If the reducible representation for the net
mobility contains mechanisms that preserve full rotational symmetry ($A$
modes), these are equiauxetic.  In addition, for $n=6$, mechanisms that halve
rotational symmetry ($B$ modes) are also equiauxetic.}

\begin{document}

\bibliographystyle{eplbib}

\maketitle

\section{Introduction} Auxetic materials (auxetics) have the property that when
stretched in one direction they expand in a perpendicular direction, that is,
have a negative Poisson's ratio.  Their proposed uses include applications in
medical, safety and sensing areas \cite{EvansAlderson:2000, YangEtAl:2004,
Liu:2006, AldersonAlderson:2007, Alderson:2011}.  Auxetic deformations are
closely related to dilatancy in granular matter
\cite{KablaSenden:2009,ZhaoSidleSwinneySchroeter:2012}, negative normal stress
in e.g.\ biopolymers \cite{Janmey:2007fv} and the negative Poisson's ratio
observed in crumpled crystalline surfaces and membranes
\cite{BoalSeifertShillcock:1993,FalcioniBowickGuitterThorleifsson:1997}.
Auxetic properties are also known to affect physical properties, such as phonon
dispersion and wave propagation or attenuation
\cite{Sparavigna:2007,Koenders:2009,RuzzeneScarpa:2005}. The focus of many
theoretical treatments of auxeticity is the identification of mechanisms at the
microscopic level that are able to account for the macroscopic behaviour of
auxetic materials and meta-materials.  These typically involve modelling of a
structure in terms of a system of hinged rotating rigid polygons
\cite{grima00,GrimaEtAl:2005,Milton:2012} or, in a complementary approach, as
an infinite bar-and-joint framework.  The bar-and-joint model has been used to
compile a catalogue of the auxetic properties of periodic tessellations of the
plane \cite{Mitschke:2009, MitschkeEtAl:2011, MitschkeEtAl:2013}.  One striking
observation that emerges from examination of this catalogue is concerned with
the functional form of the Poisson's ratio for different 2D auxetic frameworks.
There are two distinct patterns.  In many cases, Poisson's ratio, $\nu$, the
negative of the ratio of transverse to longitudinal strain, is a function of
the amount of strain.  However, in some cases $\nu$ is constant and is equal to
$-1$ for all values of strain, the unit cell changes in size but not shape, and
the auxetic behaviour for displacement along this mode is isotropic
($equiauxetic$).  A material for which Poisson's ratio attains its limiting
value of $-1$ has also been called \emph{maximally auxetic}
\cite{SunEtAl:2012}.  Many examples of what we are calling equiauxetic
behaviour have been described \cite{Milton:1992,GrimaEtAl:2000, GrimaEtAl:2008,
GrimaEtAl:2010, GrimaEtAl:2012}.  From the catalogue \cite{Mitschke:2009}, it
is also clear that equiauxetic behaviour is associated with frameworks that
have certain symmetries.  The catalogue contains examples of equiauxetic
frameworks with $p6mm$ symmetry at equilibrium, where the displaced structure
retains $p6mm$, $p6$ or $p31m$ symmetry.  Other examples have symmetry $p4mm$,
where the displaced structure retains $p4mm$ or $p4$ symmetry.

The aim of the present paper is to give a general explanation for these
observations, and to derive a symmetry-based sufficient criterion
for equiauxetic behaviour of a 2D framework.  The treatment is based
on Maxwell counting for bar-and-joint and body-and-joint frameworks, extended
to take symmetry into account \cite{fow00}, as recently adapted to cover
periodic frameworks in both 2D and 3D \cite{gf2013}.

\section{A symmetry basis for equiauxetic behaviour} The qualitative idea
needed to explain the proposed connection between symmetry and equiauxetic
behaviour in 2D relies on the fact that the only affine deformation of a
continuous body in 2D that has rotational symmetry of order greater than $2$ is
a uniform expansion/contraction.  Consider any deformation mode of the unit
cell that preserves a rotation axis of order $3$ or more.  This deformation
must be associated with equal strain in all directions, implying a Poisson's
ratio of $-1$ for the mode in question, which hence is equiauxetic.

What can be said about the symmetry of such a deformation mode?  In the
language of point groups, characters and representations, familiar from
Chemistry \cite{Bishop:1973}, the mode must belong to an irreducible
representation that has character $+1$ under proper rotations $(C_n)^q$ through
angles $2\pi q /n$.  In fact, this also implies that the mode must be
\emph{non-degenerate}, as the kernels and co-kernels for degenerate modes of
groups $C_{pv}$ ($p=6,4$) preserve at most a $C_2$ rotational axis (see
\cite{McDowell:1965}). This means all distortions within the space of
doubly-degenerate $E$-type vibrations destroy any $C_6$, $C_4$, $C_3$
rotational symmetry.

Candidates for equiauxetic modes are therefore limited to those belonging to
irreducible representations of $A$-type (those having character $+1$ under the
principal rotation $C_n$ for $n \geq 3$), and $B$-type (those having character
$-1$ under the principal rotation $C_n$ but $+1$ under $\left(C_n\right)^2$.
The plane groups (and the point groups isomorphic with their corresponding
factor groups) that can support equiauxetic modes are therefore limited to:
$p6mm$ ($C_{6v}$), $p6$ ($C_6$) ($A$ and $B$ modes); $p4mm$ ($C_{4v}$), $p4gm$
($C_{4v}$), $p4$ ($C_4$), $p3m1$ ($C_{3v}$), $p31m$ ($C_{3v}$), $p3$ ($C_3$)
($A$ modes). Thus, in the ordering used in the International Tables
\cite{inttables06}, the `auxetic plane groups' are numbered $10$ to $17$.

In the following sections, we show how modes of the required types can be
identified for microstructured cellular materials, which are appropriately
modelled as bar-and-joint \cite{Gibson:1997wm, DeshpandeFleck:2001}  or body-and-joint
frameworks.

\section{Symmetry and Mobility of Periodic Bar-and-Joint Frameworks in 2D} It
has long been recognised that counting arguments can give powerful conditions
for rigidity/mobility of structures.  The Calladine extension \cite{cal78} of
Maxwell's rule \cite{maxwell1864} gives the net mobility ($m-s$) of a \emph{finite}
bar-and-joint 2D framework as
\begin{equation}
  m-s = 2j -b -3
  \label{eq:maxwell}
\end{equation}
where $m$ is the number of mechanisms, $s$ is the number of states of
self-stress of a pin-jointed framework with $j$ joints and $b$ bars, and the
constant term accounts for rigid-body motions.  The symmetry-extended
equivalent of (\ref{eq:maxwell}) is \cite{fow00}
\begin{equation}
  \Gamma(m)-\Gamma(s) = \Gamma(j) \times \Gamma(T_x,T_y) - \Gamma(b) -
  \Gamma(T_x,T_y) - \Gamma(R_z) \label{eq:symmetrymaxwell}
\end{equation} where
each representation $\Gamma(\text{object})$ describes the symmetry of a set of
objects (which may be joints, rigid elements, points, vectors or other local
structural or dynamical motifs) in the relevant point group of the structure.
$\Gamma(\text{object})$ collects the \emph{characters}
$\chi_{\text{object}}(S)$ of sets of objects, i.e., for each symmetry operation
$S$, $\chi_{\text{object}}(S)$ is the trace of the matrix that relates the set
before and after the application of $S$.  The rigid-body terms,
$\Gamma(T_x,T_y)$ and $\Gamma(R_z)$ are the representations of the
in-plane translations and the in-plane rotation, respectively.  For further
details see \cite{Bishop:1973}.

\begin{figure}
    \begin{center}\begin{tabular}{rrrrr|rrrr}
        & \multicolumn{4}{p{2.5cm}|}{\centering Frame (a)} &
                \multicolumn{4}{|p{2.5cm}}{\centering Frame (b)} \\
        & \multicolumn{4}{p{2.5cm}|}{\centering Group $C_2$} &
                \multicolumn{4}{|p{2.5cm}}{\centering Group $C_s$} \\
                %& \multicolumn{4}{c|}{Symmetry group $C_2$} & \multicolumn{4}{|c}{Symmetry group $C_s$} \\
                & \multicolumn{4}{c|}{\includegraphics[scale=0.225]{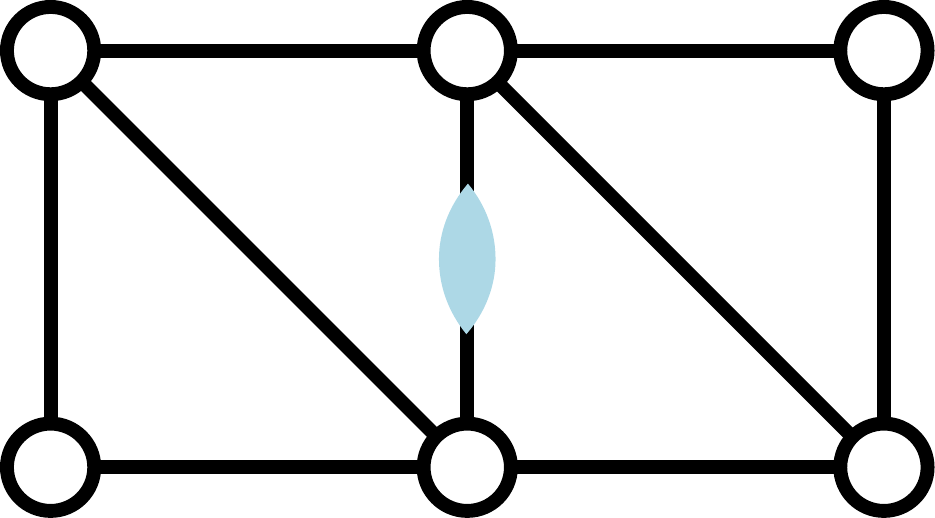}} &
                        \multicolumn{4}{|c}{\includegraphics[scale=0.225]{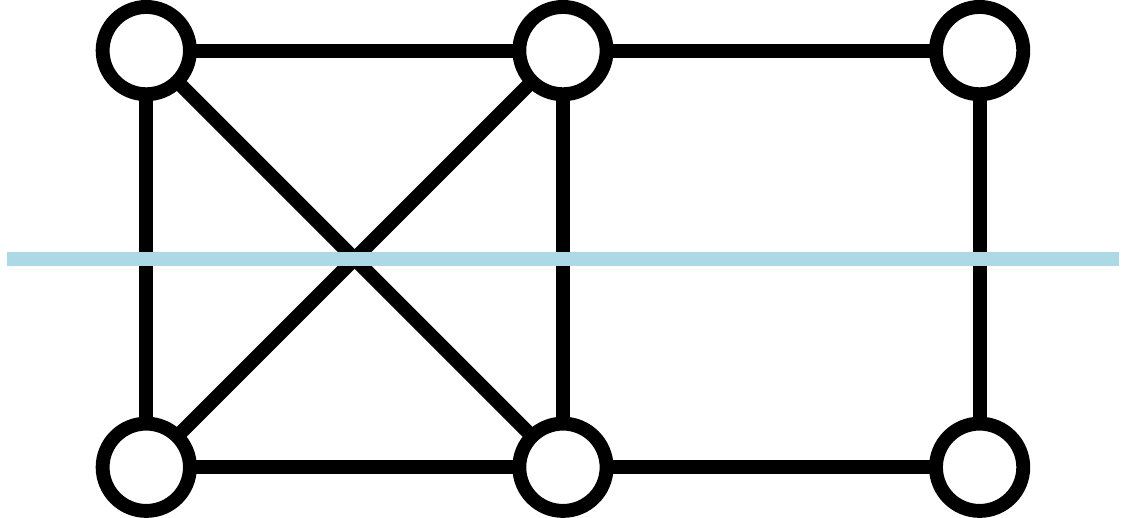}} \\

                        %\cline{2-9}
                        %& & & & & & & & \\[-12pt]
                        & \hspace{.3em} & $E$   & $C_2$ & \hspace{.5em} & \hspace{.2em} & $E$   & $\sigma$      & \hspace{0.6em} \\

                        $\Gamma(j)$                             & & $6$         & $0$   & & & $6$       & $0$   &\\
                        $\times \Gamma(T_x,T_y)$        & & $2$         & $0$   & & & $2$       & $0$   &\\[1pt]\cline{1-9}
                        & & & & & & & & \\[-12pt]
                        $=$                                             & & $12$        & $0$   & & & $12$      & $0$   &\\[6pt]
                        $- \Gamma(b)$                           & & $-9$        & $-1$  & & & $-9$      & $-3$  &\\
                        $-\Gamma(T_x,T_y)$                      & & $-2$        & $2$   & & & $-2$      & $0$   &\\
                        $-\Gamma(R_z)$                          & & $-1$        & $-1$  & & & $-1$      & $1$   &\\[1pt]\cline{1-9}
                        & & & & & & & & \\[-12pt]
                        $=\Gamma(m) - \Gamma(s)$        & & $0$         & $0$   & & & $0$       & $-2$  &
                      \end{tabular}\end{center}
  \caption{Two bar-and-joint frameworks. Both (a) and (b) comply with the
    Maxwell rule (\ref{eq:maxwell}), as expressed in the columns showing how
    the different representations behave under the trivial identity operation $E$;
    in each case the total character given in the final line of the table is
    zero.  The symmetry-extended equation (\ref{eq:symmetrymaxwell}) confirms
    the zero count for (a), where the extra symmetry operation is $C_2$ and
    gives no indication of mechanisms or states of self-stress.
    In case (b) the fact that the total character under the
    $\sigma$-reflection operation is $-2$ shows that there is at least one
    mechanism and one state of self-stress. These can be identified as the
    expected symmetry-breaking mechanism and a totally symmetric state of self-stress.}
  \label{fig:simpleframe}
\end{figure}

The terms on the RHS of (\ref{eq:symmetrymaxwell}) describe respectively,
the two-dimensional freedoms of the joints, the length constraints enforced
by the bars, and the removal of the rigid-body translations and rotation.
Each is a generalisation of the corresponding count in (\ref{eq:maxwell}).
Fig.~\ref{fig:simpleframe} shows an example of a simple system for which
symmetry provides extra information not available from (\ref{eq:maxwell}).

The scalar counting rule (\ref{eq:maxwell}) is the character of the full
symmetry equation under the identity operation;  it can be extended to
\emph{periodic} structures.  When proper account is taken of the
allowed degrees of freedom of the lattice \cite{gh03} (stretches and shear
motions) the form of (\ref{eq:maxwell}) appropriate to a periodic system in 2D
is \cite{gf2013}
\begin{equation}
m-s = 2j -b +1.
\label{eq:maxwellperiodic}
\end{equation}
Extending (\ref{eq:maxwellperiodic}) to include periodic symmetry gives \cite{gf2013}:
\begin{equation}
\Gamma(m)-\Gamma(s) = \Gamma(j) \times \Gamma(T_x,T_y) - \Gamma(b) + \Gamma_a
\label{eq:symmperiodic}
\end{equation}
where
\begin{equation}
    \Gamma_a = \Gamma(T_x,T_y) \times \Gamma(T_x,T_y) - \Gamma(T_x,T_y) - \Gamma(R_z),
\label{eq:a}
\end{equation}
and all representations $\Gamma$ are to be calculated in the crystallographic
point group isomorphic with the factor group of the full plane group.  The
representation $\Gamma_a$ accounts for the difference in
symmetry between the three possible deformations of the unit cell and the two
allowed rigid-body displacements.
\begin{figure}
  \centering
    \begin{subfigure}{0.485\columnwidth}
    \centering
    \includegraphics[scale=0.85]{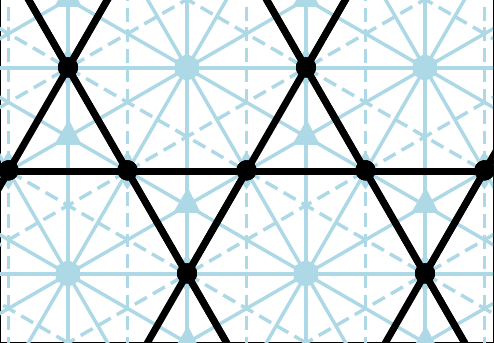}
    \caption{}
    \label{fig:kagome-p6mm}
  \end{subfigure}
  \begin{subfigure}{0.485\columnwidth}
    \centering
    \includegraphics[scale=0.85]{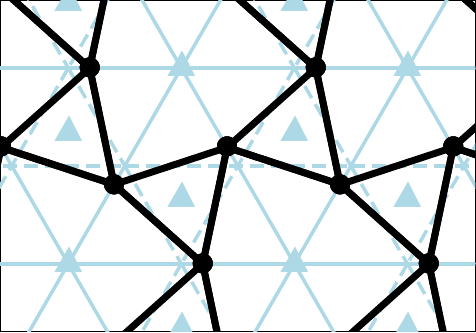}
    \caption{}
    \label{fig:kagome-p31m}
  \end{subfigure}
  \caption{(a) Kagome framework shown with $p6mm$
  symmetry elements; (b) symmetry-detected $B_2$
mechanism, retaining subgroup $p31m$.} \label{fig:kagome}
\end{figure}

A convenient format for the application of this equation to a periodic
framework with some point symmetry is the following tabulation, given here for
the kagome lattice (Fig.~\ref{fig:kagome}) which has factor group ${\mathcal P}
\cong C_{6v}$ in the notation of  \cite{BurnsGlazer:1978} which makes clear the
connection between the point group and the crystallographic plane group, in
this case $p6mm$.  \\[0.5em]
  \begin{tabular}{*{8}{r}}
    %\multicolumn{1}{l}{\footnotesize{Kagome:}}  &&&&&&&\\[1pt]\cline{1-7}\\[-12pt]
    $C_{6v}$             & $E$ & $2C_6$ & $2C_3$& $C_2$ & $3\sigma_v$& $3\sigma_d$ & \\[1pt]\cline{1-7}\\[-12pt]
$\Gamma(j)$               & 3   & 0      & 0     & 3     & 1          & 1           & \\
$\times \Gamma(T_x,T_y)$ & 2   & 1      & $-1$    & $-2$    & 0          & 0           & \\[1pt]\cline{1-7}\\[-12pt]
$=$                   & 6   & 0      & 0     & $-6$    & 0          & 0               & \\[6pt]
$- \Gamma(b)$             & $-6$  & 0      & 0     & 0     & $-2$         & 0           & \\
$+ \Gamma_a$              & 1   & $-1$     & 1     & 5     & 1          & 1           & \\[1pt]\cline{1-7}\\[-12pt]
$\Gamma(m) - \Gamma(s)$   & 1   & $-1$     & 1     & $-1$    & $-1$         & 1
\end{tabular}  \\[0.5em]
The result is that $\Gamma(m) - \Gamma(s) = B_2$.  The conclusion is that
$\Gamma(m)$ contains a mechanism of $B_2$ symmetry, and there are no
symmetry-detectable states of self-stress.  The detected mechanism is the well
known periodic collapse mode for the kagome lattice \cite{kap09}, where
alternate triangles rotate in opposite senses, as seen in
Fig.~\ref{fig:kagome}(b). This in fact, is an auxetic mode.

As an illustration of the additional qualitative information provided by the
tabulation, consider the $C_2$ column in this table. The third entry shows that
all six freedoms of the joints are reversed by the $C_2$ operation; all three
points are unshifted by the operation, but it reverses the attached $x$ and $y$
vectors.  The entry for $\Gamma(b)$ is zero as all bars are shifted by the
operation. The value of $5$ for $\Gamma_a$ arises because the three allowed
deformations of the unit cell are symmetric under $C_2$ but the two rigid-body
motions are antisymmetric ($5=+3-(-2)$). The final total of $-1$ for the
character of $\Gamma(m) - \Gamma(s)$ under $C_2$ tells us that the mechanism
that we have detected by using (\ref{eq:symmperiodic}) breaks this symmetry of
the lattice.

It should be noted that although equations such as (\ref{eq:symmetrymaxwell})
and (\ref{eq:symmperiodic}) can be powerful in revealing more detail than would
be accessible through scalar counting alone, they do have an important
limitation in that they necessarily yield only the representation of the
relative mobility.  If a structure has mechanisms that are equisymmetric with
states of self-stress, there will be no evidence for them in $\Gamma(m) -
\Gamma(s)$.  By definition, then, the mechanisms revealed by
(\ref{eq:symmperiodic}) are \emph{symmetry-detectable}.  The fact that
symmetry-detectable mechanisms may become undetectable on descent in symmetry
can itself be used to diagnose finite \emph{vs.} infinitesimal mechanisms.  The
scalar counterpart of detectability is that relationships such as
(\ref{eq:maxwell}) and (\ref{eq:maxwellperiodic}) give only a lower bound on
the number of mechanisms.

\section{A criterion for equiauxetic behaviour} Combination of the reasoning
about rotational symmetry of auxetic modes with the symmetry-extended mobility
rule (\ref{eq:symmperiodic}) leads to the following statement.

\emph{Auxeticity Criterion}: A periodic 2D system with plane group $\cal{G}$
and factor group $\cal P = \cal{G}/\mathcal{T}$ has symmetry-detectable
equiauxetic behaviour if and only if
\begin{enumerate}
    \item $\cal P$ is isomorphic to a point group from the list $C_{6v}$,
      $C_{6}$, $C_{4v}$, $C_{4}$, $C_{3v}$, $C_{3}$, and
    \item the reducible representation $\Gamma(m)-\Gamma(s)$ contains one or
      more copies of an \emph{auxetic} irreducible representation.  The auxetic
      irreducible representations are: $A_1$, $A_2$, $B_1$, $B_2$ in $C_{6v}$;
      $A$, $B$ in $C_{6}$; $A_1$, $A_2$ in $C_{4v}$; $A$ in $C_{4}$; $A_1$,
      $A_2$ in $C_{3v}$; $A$ in $C_{3}$. ($\Gamma(m)-\Gamma(s)$ can be computed
      according to \eqref{eq:symmperiodic} for bar-and-joint frameworks, and
      \eqref{eq:symbmobsimpper} for body-and-joint structures)
\end{enumerate}

For practical calculations it is useful to note the composition of $\Gamma_a$
within the relevant point groups:
\begin{equation}
  \Gamma_a=\left\{%
\begin{array}{@{}llll@{\qquad}l}
  A_1 &-E_1 &+E_2 &&(C_{6v})\\
  A   &-E_1 &+E_2 &&(C_{6})\\
  A_1 &+B_1 &+B_2 &-E&(C_{4v})\\
  A   &+2B  &-E&&(C_{4})\\
  A_1 &&&&(C_{3v})\\
  A&&&&(C_{3})
\end{array}%
\right.
\label{eq:gamma_a}
\end{equation}

An alternative way of reaching the same conclusion, in the manner of
\cite{ConnellyEtAl:2009}, is to calculate the number of times each auxetic
representation occurs in the (reducible) representation $\Gamma(m)-\Gamma(s)$,
i.e., to find the coefficients $n(\Gamma_i)$ in the expansion
\begin{equation}
  \Gamma(m)-\Gamma(s) = \sum_i n(\Gamma_i) \Gamma_i
\end{equation}
where $i$ runs over the irreducible representations of the group.  The
coefficients $n(\Gamma_i)$ can be found by using well known projection
techniques \cite{Bishop:1973}.  It is straightforward to show that the counts
for the auxetic representations are:\\ $C_{6v} \cong p6mm/\mathcal{T}$ and
$C_{4v} \cong p4mm/\mathcal{T} \cong p4gm/\mathcal{T}$:
\begin{equation}\begin{split}
    n(A_1) &= 2j_1 +j_{m} - (b_1 + b_{m} + b_{2mm}) +1 \qquad\\
      n(A_2) &= 2j_1 +j_{m} - \;b_1 \\
\end{split}\label{eq:counts1}\end{equation}
$C_{6v} \cong p6mm/\mathcal{T}$:
\begin{equation}\begin{split}
        n(B_1) &= 2j_1 +j_{m} + j_{2mm} - (b_1 +b_{.m.})\\
          n(B_2) &= 2j_1 +j_{m} + j_{2mm} - (b_1 +b_{..m})
\end{split}\end{equation}
$C_{3v} \cong p31m/\mathcal{T} \cong p3m1/\mathcal{T}$:
\begin{equation}\begin{split}
      n(A_1) &= 2j_1+j_{m} - (b_1 +b_{m}) +1\\
        n(A_2) &= 2j_1+j_{m} - \;b_1
\end{split}\end{equation}
$C_6=p6/\mathcal{T}$ and $C_4=p4/\mathcal{T}$:
\begin{equation}\begin{split}
  n(A) &= 2j_1 -(b_1+b_{2}) +1\\
\end{split}\end{equation}
$C_6=p6/\mathcal{T}$:
\begin{equation}\begin{split}
    n(B) &= 2(j_1+j_{2})  - b_1
\end{split}\end{equation}
$C_3=p3/\mathcal{T}$:
\begin{equation}
  n(A) = 2j_1-b_1+1
  \label{eq:counts2}
\end{equation}
Here, the numbers of joints and bars are those in the asymmetric unit of the
plane group, subscripted with their (oriented) site symmetry symbol:  $j_1$ and
$b_1$ are in general position;  $j_2$ and $b_2$ are in special positions having
2-fold rotational symmetry; $j_m$ and $b_m$ are on mirror lines; and $j_{2mm}$
and $b_{2mm}$  are in sites of symmetry $C_{2v} \equiv 2mm$. Dots are
used, in the fashion of \cite{inttables06}, where necessary to
distinguish settings of the mirror line associated with $C_s \equiv m$.

\section{Symmetry and Mobility of Periodic Body-and-Joint Structures in 2D}
The scalar and symmetry-extended counting rules developed so far apply to the
bar-and-joint model of a repetitive structure.  An alternative model is often
employed to describe solid-state materials, where the structure is considered
as consisting of rigid units connected through flexible joints
\cite{giddy93}. For some systems this model may be more appropriate than the
use of bars and joints, as will appear below in our treatment of the
`TS-wheels' tiling \cite{Mitschke:2009,MitschkeEtAl:2013}. A symmetry-extended
counting rule for the mobility of periodic body-and-joint structures has been
derived \cite{gf2013} and the auxetic criterion discussed in the previous
section applies equally to this model.

The model for the 2D case considers the relative degrees of freedom of a
repetitive mechanical linkage consisting of unit cells containing $n$ bodies
connected by $g$ joints, where in this case each joint permits a single
relative rotational freedom.  The scalar counting rule is
\begin{equation}
    m-s = 3n - 2g + 1
    \label{eq:bhperiod}
\end{equation}
where the RHS expresses the facts that each body has three degrees of freedom
in the plane, each hinge constrains two degrees of freedom, and the repetitive
nature of the system gives the additional single freedom, as discussed above in
relation to (\ref{eq:maxwellperiodic}).  The symmetry extension of
(\ref{eq:bhperiod}) is cast in terms of properties of the `contact polyhedron'
$C$, which is in fact an infinite object, but one for which we need only
consider a repeating unit cell.  The vertices of $C$ represent the rigid
elements, and the edges joints.  As described in \cite{gf2013}, the
symmetry-extended counting rule for this case is
\begin{multline}
  \Gamma(m)-\Gamma(s) = \Gamma(v,C) \times \left(\Gamma(T_x,T_y) +
  \Gamma(R_z)\right) -\\ \Gamma_\|(e,C) \times \Gamma(T_x,T_y) + \Gamma_a
  \label{eq:symbmobsimpper}
\end{multline}
where $\Gamma_a$ is the same representation as defined in \eqref{eq:a}.

With the aid of (\ref{eq:symmperiodic}) and  (\ref{eq:symbmobsimpper}) we
now have the conditions for symmetry-detectable equiauxetic mechanisms
in both bar-and-joint and body-and-joint
structures in 2D.

\section{Examples}
In the following, we discuss a number of example frameworks drawn from
\cite{Mitschke:2009} chosen to illustrate points of particular interest.  The
names for the various frameworks are those used in the catalogue, and are
supplemented with the symbol for the tiling taken from \cite{gs87}.  For each
example the plane group, the point group, and the net mobility are listed.  The
examples range from overconstrained, with $m-s$ negative, to underconstrained,
with $m-s$ positive.  In each case a picture and a tabular calculation are
given.  Each picture shows the framework (in bold) and symmetry elements
according to the conventions of the International Tables \cite{inttables06},
together with translation vectors delineating the unit cell.

\subsection{Kagome}
This is tiling $(3.6.3.6)$, $p6mm$, $6mm(C_{6v})$, $m-s=1$.  The tabular form of
the symmetry-adapted mobility calculation was given earlier.  With the chosen
unit cell (Fig.~\ref{fig:kagome}), scalar counting gives $m-s=1$, implying the
existence of at least one mechanism.  The symmetry calculation reveals that
this mechanism is equiauxetic (and of $B_2$ symmetry).  Numerical calculations
show that the mechanism is unique for this smallest choice of unit cell.  As
Fig.~\ref{fig:kagome}(b) shows, the plane group for the deformed configuration
is $p31m$, describing a lattice in which each triangle rotates in an opposite
sense to its neighbours.

\subsection{TS-wheels}
\begin{figure}
  \centering
  {\begin{subfigure}{0.485\columnwidth}
    \centering
    \includegraphics[scale=0.85]{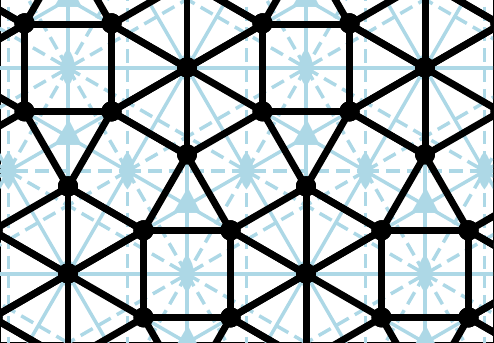}
    \caption{}
    \label{fig:2uni5_p6mm}
  \end{subfigure}}
  \begin{subfigure}{0.485\columnwidth}
    \centering
    \includegraphics[scale=0.85]{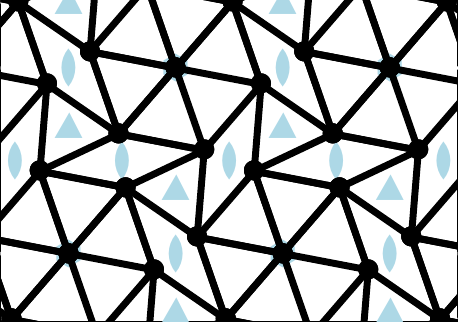}
    \caption{}
    \label{fig:2uni5_p6_def}
  \end{subfigure}\\
  \begin{subfigure}{0.485\columnwidth}
      \centering
      \includegraphics[scale=0.85]{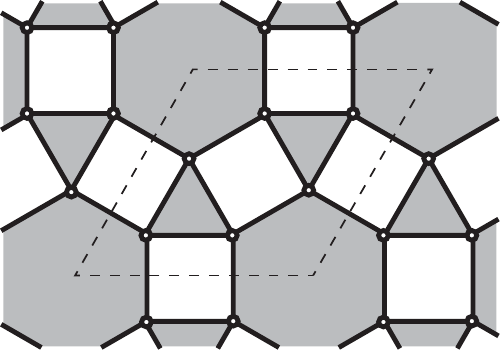}
      \caption{}
      \label{fig:2uni5_bj}
    \end{subfigure}
  \begin{subfigure}{0.485\columnwidth}
      \centering
      \includegraphics[scale=0.85]{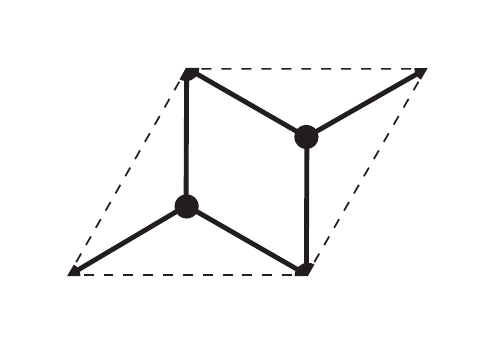}
      \caption{}
      \label{fig:2uni5_cp}
    \end{subfigure}
  \caption{(a) The TS-wheels tiling shown with $p6mm$ symmetry
  elements; (b) $A_2$ mechanism retaining $p6$ symmetry; (c)
  TS-wheels tiling shown as a body-joint model; (d) the contact
polyhedron for the body-joint model.} \label{fig:2uni5}
\end{figure}
This overconstrained framework is the 2-uniform tiling $(3^6;3^2.4.3.4)$,
$p6mm$, $6mm(C_{6v})$, $m-s=-3$
(Fig.~\ref{fig:2uni5}(\subref{fig:2uni5_p6mm})). Simple counting indicates only
the existence of three states of self-stress. Counting with symmetry, as in the
tabular calculation shown below, detects a mechanism of $A_2$ symmetry, which
corresponds to concerted rotation of the hexagonal `wheels', with simultaneous
collapse of the square motifs to flattened rhombi.\\[0.5em]
  \begin{tabular}{*{7}{r}}
    $C_{6v}$                 & $E$   & $2C_6$ & $2C_3$ & $C_2$ & $3\sigma_v$ & $3\sigma_d$ \\[1pt]
    \cline{1-7}\\[-12pt]
    $\Gamma(j)$              & 7     & 1      & 1      & 1     & 3           & 1           \\
    $\times \Gamma(T_x,T_y)$ & 2     & 1      & $-1$   & $-2$  & 0           & 0           \\[1pt]
    \cline{1-7}\\[-12pt]
    $=$                      & 14    & 1      & $-1$   & $-2$  & 0           & 0           \\[6pt]
    $- \Gamma(b)$            & $-18$ & 0      & 0      & 0     & $-4$        & $-2$        \\
    $+ \Gamma_a$             & 1     & $-1$   & 1      & 5     & 1           & 1           \\[1pt]
    \cline{1-7}\\[-12pt]
    $\Gamma(m) - \Gamma(s)$  & $-3$  & 0      & 0      & 3     & $-3$        & $-1$        \\[4pt]
  \end{tabular}\\[0.5em]
Thus $\Gamma(m) - \Gamma(s) = A_2 - A_1 -B_1 -E_1$, implying at least one mechanism,
of $A_2$ symmetry, and four states of self stress,
of symmetries $A_1$, $B_1$ and $E_1$.

Numerical calculation \cite{MitschkeEtAl:2013} shows that the symmetry-detected
mechanism is the sole mechanism for this choice of unit cell and remains so for
larger cells corresponding to $n \times n$ multiples, for at least
$n \le 6$.  By its symmetry, this mechanism is equiauxetic.  Experimental
results for a constructed cellular metamaterial with rigid joints, and
finite-element simulations of the bar-and-joint framework confirm this
prediction of symmetry analysis \cite{MitschkeEtAl:2011}.

The four states of self-stress predicted by symmetry have respective
representations $A_1$, $B_1$ and $E_1$; the mechanism would therefore become
equisymmetric with the first state of self-stress if followed down from $p6mm$
to $p6$, satisfying a necessary condition for `blocking' of the mechanism,
implying that it might be infinitesimal in character rather than finite
\cite{kangwai1999a,GF07}.  However, it is easy to see here that the relevent
state of self-stress is localised within the hexagonal wheel and cannot block
the rotational mechanism.  The $A_2$-symmetric mechanism will continue all the
way to the collapsed state where all quadrilaterals have flattened out,
yielding a degenerate version of the kagome framework with superimposed bars.

The irrelevance of the `internal' state of self-stress to blocking the
mechanism is readily apparent in the alternative body-and-joint model of
2-uniform 5 shown in Fig.~\ref{fig:2uni5}(\subref{fig:2uni5_bj}).
Fig.~\ref{fig:2uni5}(\subref{fig:2uni5_cp}) shows the contact polyhedron, with
vertices corresponding to rigid triangular bodies.  A tabular calculation
according to (\ref{eq:symbmobsimpper}) is shown below.
The result $\Gamma(m) - \Gamma(s) = A_2 - B_1 - E_1$ reveals the same
symmetry-detected $A_2$ mechanism, but now only three states of self-stress,
and in particular no state of self-stress that is totally symmetric in $p6mm$.
The finite nature of the equiauxetic $A_2$ mode is therefore apparent from this
more physically insightful choice of model structure.\\[0.5em]
  \begin{tabular}{*{7}{r}}
    $C_{6v}$                     & $E$   & $2C_6$ & $2C_3$ & $C_2$ & $3\sigma_v$ & $3\sigma_d$  \\[1pt]\cline{1-7}\\[-12pt]
    $\Gamma(v,C)$                & 3     & 1      & 3      & 1     & 3           & 1            \\
    $\times \Gamma(T_x,T_y,R_z)$ & 3     & 2      & 0      & $-1$  & $-1$        & $-1$         \\[1pt]\cline{1-7}\\[-12pt]
    $=$                          & 9     & 2      & 0      & $-1$  & $-3$        & $-1$         \\[6pt]
    $- \Gamma_\parallel(e,C)$    & $-6$  & 0      & 0      & 0     & $-2$        & 0            \\
    $\times \Gamma(T_x,T_y)$     & 2     & 1      & $-1$   & $-2$  & 0           & 0            \\[1pt]\cline{1-7}\\[-12pt]
    $=$                          & $-12$ & 0      & 0      & 0     & 0           & 0            \\[6pt]
    $+ \Gamma_a$                 & 1     & $-1$   & 1      & 5     & 1           & 1            \\[1pt]\cline{1-7}\\[-12pt]
    $\Gamma(m) - \Gamma(s)$      & $-2$  & 1      & 1      & 4     & $-2$        & 0
  \end{tabular}\\[0.5em]

\begin{figure*}
    \centering
      \begin{subfigure}{0.32\linewidth}
        \centering
        \includegraphics[scale=1.0]{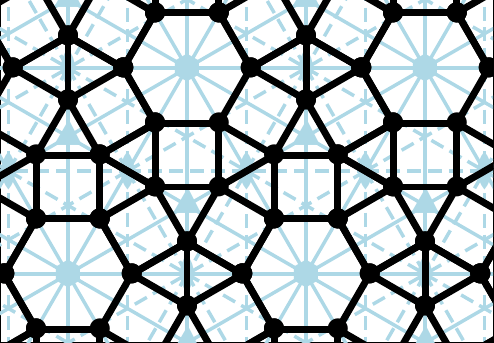}
        \caption{}
        \label{fig:2uni14_p6mm}
      \end{subfigure}
      \begin{subfigure}{0.32\linewidth}
        \centering
        \includegraphics[scale=1.0]{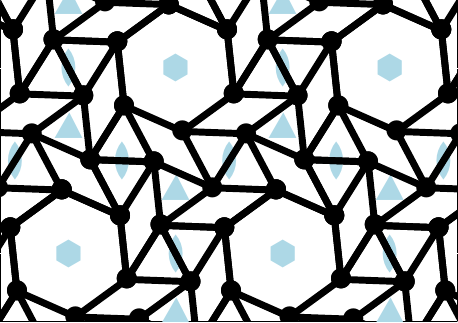}
        \caption{}
        \label{fig:2uni14_p6_def}
      \end{subfigure}
      \begin{subfigure}{0.32\linewidth}
        \centering
        \includegraphics[scale=1.0]{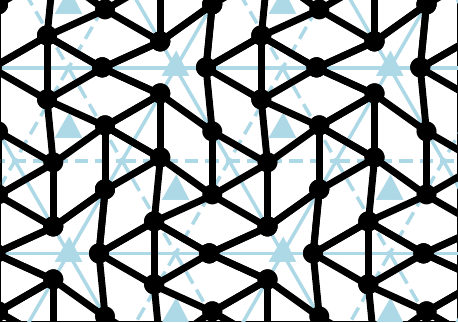}
        \caption{}
        \label{fig:2uni14_p31m_def}
      \end{subfigure}
      \caption{(a) 2-uniform tiling
      ${(3^2.4.3.4;3.4.6.4)}$ shown with $p6mm$ symmetry elements;
      (b) $A_2(p6)$ mechanism (unit cell contracted by
      a factor of $0.1$); (c) $B_2(p31m)$
    mechanism.} \label{fig:2uni14}
\end{figure*}

\subsection{2-uniform tiling 14}
This tiling has the description ${(3^2.4.3.4;3.4.6.4)}$, $p6mm$, $6mm(C_{6v})$,
$m-s=-2$.  The tiling is illustrated in Fig.~\ref{fig:2uni14}(a) and is
included as an example of the possibility of multiple auxetic
pathways.\\[0.5em]
\begin{tabular}{*{8}{r}}
  $C_{6v}$                 & $E$   & $2C_6$ & $2C_3$ & $C_2$ & $3\sigma_v$ & $3\sigma_d$ & \\[1pt]\cline{1-7}\\[-12pt]
  $\Gamma(j)$              & 12    & 0      & 0      & 0     & 2           & 2           & \\
  $\times \Gamma(T_x,T_y)$ & 2     & 1      & $-1$   & $-2$  & 0           & 0           & \\[1pt]\cline{1-7}\\[-12pt]
  $=$                      & 24    & 0      & 0      & 0     & 0           & 0           & \\[6pt]
  $- \Gamma(b)$            & $-27$ & 0      & 0      & $-3$  & $-5$        & $-1$        & \\
  $+ \Gamma_a$             & 1     & $-1$   & 1      & 5     & 1           & 1           & \\[1pt]\cline{1-7}\\[-12pt]
  $\Gamma(m) - \Gamma(s)$  & $-2$  & $-1$   & 1      & 2     & $-4$        & 0           &
\end{tabular}\\[0.5em]
The final result is $\Gamma(m) - \Gamma(s) = A_2 + B_2 - A_1 - B_1 - E_1$,
leading to a resolution of the scalar count $m-s$ into a balance of two
mechanisms belonging to distinct representations  $A_2$ and $B_2$ and three
states of self-stress belonging to $A_1 + B_2 +E$.  The two mechanisms
correspond to distinct symmetry-reducing pathways, shown in
Fig.~\ref{fig:2uni14}(\subref{fig:2uni14_p6_def},\subref{fig:2uni14_p31m_def}),
leading to different subgroups of $p6mm$.  The $A_2$ pathway retains $p2$
symmetry, and the $B_2$ pathway retains centrosymmetric mirror symmetry $cm$.
Both mechanisms are found to be finite \cite{MitschkeEtAl:2011}. The symmetry
results indicate potential blocking of the mechanism by an equisymmetric state
of self stress, but this does not materialise.

\section{Conclusions}
In this work, symmetry considerations have been used to give a basis for
understanding generic isotropic auxetic (equiauxetic) mechanisms of 2D
materials and meta-materials with sufficiently high symmetry. Explicit criteria
for the detection and characterisation of such mechanisms have been given here
only for bar-and-joint and body-and-joint models of frameworks, but could be
extended to any model with a well defined notion of counting degrees of
freedom. For symmetric systems, these criteria are often more informative than
purely combinatorial approaches.  When a system has a unique equiauxetic mode,
it will exhibit auxetic behaviour; in the general case, the equiauxetic mode
may be accompanied by other modes which may provide alternative non-equiauxetic
deformations.

The theory developed here for 2D materials has a 3D counterpart and the
necessary symmetry-extended mobility criteria have been given in \cite{gf2013}.
In 3D the only affine deformation that preserves cubic symmetry is uniform
expansion/contraction.  The problem of finding 3D systems with unique
equiauxetic modes is challenging.

A direction of future exploration beyond the present periodic approach is for
materials where local symmetry coexists with long-range disorder.  Elastomeric
polypropylene films \cite{Magerle:2011} are of this type, and the Poisson's
ratio and rigidity of random network structures is of interest in several other
physical systems \cite{Bhaskar:2009,DelaneyWeaireHutzler:2005}.

\acknowledgments We acknowledge financial support by the DFG through the
{Engineering of Advanced Materials Cluster of Excellence} and the research
group GPSRS under Grants SCHR1148/3-1 and ME1361/12-1.  PWF acknowledges a
Royal Society/Leverhulme Senior Research Fellowship.

\end{document}